\documentclass[aps,showpacs,manuscript,12pt]{revtex4}
\usepackage{amssymb}
\usepackage{amsmath}
\usepackage{graphicx}
\usepackage{epsfig}

\begin{document}

\title{\textbf{Non-linear charge reduction effect in strongly-coupled plasmas%
}}
\author{D. Sarmah$^{1,2,3}$, M. Tessarotto$^{1,3}$ and M. Salimullah$^{4}$}
\affiliation{$^{1}$Department of Mathematics and Informatics,\\
University of Trieste, Trieste, Italy \\
$^{2}$ International Center for Theoretical Physics, ICTP/TRIL\\
Program, Trieste, Italy \\
$^{3}$ Consortium for Magnetofluid Dynamics, Trieste, Italy\\
$^{4}$ Department of Physics, Jahangirnagar University, Savar, Dhaka,\\
Bangladesh}

\begin{abstract}
The charge reduction effect, produced by the nonlinear Debye screening of
high-Z charges occuring in strongly-coupled plasmas, is investigated. An
analytic asymptotic expression is obtained for the charge reduction factor ($%
f_{c}$) which determines the Debye-H\"{u}ckel potential generated by a
charged test particle. Its relevant parametric dependencies are analyzed and
shown to predict a strong charge reduction effect in strongly-coupled
plasmas.
\end{abstract}

\date{\today }
\maketitle

A basic aspect of plasma physics is the so-called Debye shielding of the
electrostatic potential. \ This consists in the property of plasmas  (or
electrolytes \cite{Debye 1923}), either quasi-neutral or non-neutral, to
shield the electrostatic field produced by charged particles (to be also
denoted as test particles) immersed in the same system. This result has
fundamental consequences on plasma phenomenology, since it actually limits
the range of static Coulomb interactions inside the Debye sphere, i.e., at a
distance $\rho \leq \lambda _{D}$ from the test particle$,$ $\lambda _{D}$
being the Debye length. As usual here $\lambda _{D}\equiv \left(
\sum\limits_{s}\lambda _{Ds}^{-1}\right) ^{-1},$ where the sum is carried
out on all plasma species and $\lambda _{Ds}=\sqrt{\frac{T_{s}}{4\pi
Z_{s}^{2}e^{2}N_{os}}},$ $T_{s}$ and $N_{os}$ being respectively the $s-$%
species temperature and number density (the latter defined in the absence of
test particles). In fact, when both the particles and the plasma are assumed
non-relativistic, small-amplitude, stationary (or slowly time- and
space-varying), electrostatic perturbations generated by isolated test
particles, result effectively shielded in the external domain, i.e., at
distances larger than the Debye length $\lambda _{D}$. The renewed interest
in this problem is particularly related to dusty plasmas or colloidal
suspensions \cite{Pieransky 1983,Thomas 1994} which are characterized by the
presence of a large fraction of highly charged particles (grains), i.e.,
having an electric charge $Z_{d}e$ with $\left\vert Z_{d}\right\vert \gg 1$.
It is well established that the phenomenon of Debye shielding of the
electrostatic potential generated by a slowly moving or stationary charged
test particle, when occurring in strongly coupled ionized gases, manifests
peculiar properties. These are produced, as a consequence of the nonlinear
plasma response occurring at distances smaller than the Debye length $%
\lambda _{D}$, when the electrostatic potential ($\Phi $) results locally
such that $\widehat{\Phi }\equiv \left\vert e\Phi \right\vert /T_{o}\sim 1$
(or even $\gg 1$)$,$ $\widehat{\Phi }$ being the normalized electrostatic
potential evaluated at a suitable characteristic distance $\rho _{o}$ from
the position of the test particle $\mathbf{r}(t)$ and $T_{o}\equiv T_{i}$.
The distance $\rho _{o}$ can be identified with the radius of the local
plasma sheath around a test particle. For spherically-symmetric test
particles such a domain can be identified with the spherical subset of the
configuration space, of radius $\rho _{o}\equiv \left( \frac{3}{4\pi N_{o}}%
\right) ^{1/3}$ (average interparticle distance),  $N_{o}\equiv N_{oi}$
being the density of the ion species, in which each test particle can be
considered as isolated. Ionized gases (or electrolytes) can be classified
according to the characteristic dimensionless parameters $x_{o}=\rho
_{o}/\lambda _{D}\equiv g^{1/3}$ and $\Gamma \equiv \beta /x_{o}$ (where $%
\beta =\frac{Z_{d}e^{2}}{4\pi T_{o}\lambda _{D}}$ is denoted as the
dimensionless electric charge of the test particle), $g$ and $\Gamma $ \
being respectively so-called \emph{plasma} and \emph{Coulomb coupling
parameters}. In particular, for plasmas the high-density requirement $%
x_{o}^{3}\equiv g\ll 1$ is satisfied by assumption and can be interpreted as
the condition that the total number of mutually interacting particles in a
Debye sphere (expressed by $1/g$) results $\gg 1$. Instead, the ordering of
parameter $\Gamma $ remains in principle arbitrary. Thus, the orderings $%
\Gamma \ll 1$ and $\Gamma \sim 1$ (or $\Gamma \gg 1$) correspond
respectively to so-called weakly and strongly-coupled plasmas. \ Dusty and
colloidal plasmas may be characterized \ by $Z_{d}\ $up to $10^{4}-10^{6}$
with typical grain size smaller than $10^{-4}$ $cm$ and with plasma
temperature and the ion density $T_{o}\sim 1$ eV, $N_{o}\sim 10^{8}-10^{10}$
cm$^{-3}$. In this case the\ Coulomb parameter for a negatively-charged
grain in the presence of the plasma sheath produced by hydrogen ions can
result as large as $\Gamma \cong (10^{-2}-10^{-3})Z_{d},$ while the
dimensionless radius of the ion plasma sheath can be estimated $x_{o}\cong
0.01-0.05.$ Therefore, dusty plasmas typically result strongly coupled if $%
Z_{d}\gtrsim 10^{3}$. For these plasmas it is important to be able to
evaluate the electrostatic potential generated by dust grains which is
expected to be strongly influenced by Debye shielding. It is well-known
that, in general, this phenomenon occurs provided suitable physical
assumptions are introduced. In particular, the plasma must be assumed
appropriately close to kinetic Maxwellian equilibrium, in which each
particle species is described by a Maxwellian kinetic distribution function
carrying finite fluid fields [defined respectively by the number density,
temperature and flow velocity $\left( N,T,\mathbf{V}\right) $]. In the
absence of test particles these fluid fields must be assumed slowly varying
in a suitable sense, or constant, both with respect to position ($\mathbf{r}$%
) and time ($t$). \ In this regard it is important to remark that the
appropriate treatment of the plasma sheath surrounding each test particle is
essential also for the validity of the mathematical model for the Debye
screening problem, i.e., for the existence of classical solutions of the
Debye screening problem, which do not exist when letting $x_{o}=0$ \cite%
{Tessarotto2005}. As a consequence the electrostatic potential depends
necessarily on the dimensionless parameter $x_{o}$. In particular, in the
domain $x\geq x_{o}$ the dimensionless potential $\widehat{\Phi }_{x_{o}}(x)$
generated by a dust grain of charge $Z_{d}e<0,$ fulfilling the boundary
conditions $\left. \frac{d\widehat{\Phi }_{x_{o}}(x)}{dx}\right\vert
_{x=x_{o}}=-\frac{\Gamma }{x_{o}}$ and lim$_{x\rightarrow \infty }\widehat{%
\Phi }_{x_{o}}(x)=0,$ can be proven to satisfy the \emph{integral
Debye-Poisson equation}%
\begin{equation}
\widehat{\Phi }_{x_{o}}(x)\mathbf{=}\frac{\beta }{x}-\left[ \frac{1}{x}%
\int_{x_{o}}^{x}dx^{\prime }x^{\prime 2}+\int_{x}^{\infty }dx^{\prime
}x^{\prime }\right] S(x^{\prime },x_{o}),  \label{eq.0}
\end{equation}%
yielding for $x=x_{o}$ the constraint equation%
\begin{equation}
\widehat{\Phi }_{o}=\Gamma -\int_{x_{o}}^{\infty }dx^{\prime }x^{\prime
}S(x^{\prime },x_{o}).  \label{eq.00}
\end{equation}%
Here $\widehat{\Phi }_{o}\equiv \widehat{\Phi }_{x_{o}}(x_{o}),$ while in
the case of a negatively-charged test particles in a quasi-neutral three
species (electron, ion and dust) plasma the source term $S(x,x_{o})$ reads $%
S(x,x_{o})\equiv \Theta (x-x_{o})\left[ \xi _{i}\exp \left\{ \left\vert
Z_{i}\right\vert \widehat{\Phi }_{x_{o}}(x)\right\} -\xi _{e}\exp \left\{ -%
\widehat{\Phi }_{x_{o}}(x)\right\} -\xi _{d}\exp \left\{ -\left\vert
Z_{d}\right\vert \widehat{\Phi }_{x_{o}}(x)\right\} \right] ,$\  $\Theta
(x-x_{o})$ being the Heaviside function. Moreover, $\xi _{s}$ is defined by
the ratio $\xi _{s}=\frac{1}{\left\vert Z_{s}\right\vert }\frac{\lambda
_{D}^{2}}{\lambda _{Ds}^{2}}.$ In the case in which the contribution due to
the dusty species is negligible, and moreover $\left\vert Z_{i}\right\vert
=1,$ $T_{i}=T_{e},$ the previous expression reduces to the customary value $%
S(x,x_{o})\equiv \Theta (x-x_{o})\sinh \widehat{\Phi }_{x_{o}}(x)$. Another
useful representation of Eq.(\ref{eq.0}) can also be obtained by means of
the transformation
\begin{equation}
\widehat{\Phi }_{x_{o}}(x)=\widehat{\Phi }_{o}\exp \left\{
y(x,x_{o})\right\} ,  \label{eq.0c}
\end{equation}%
and there results, by consistency with the integral DP equation (\ref{eq.0}%
), $y(x_{o},x_{o})=0$ and $\left. y^{\prime }(x,x_{o})\right\vert
_{x=x_{o}}=-\alpha $, with $\alpha =\Gamma /x_{o}\widehat{\Phi }_{o}$. This
delivers for $y(x,x_{o})$ the integral equation%
\begin{eqnarray}
&&\left. y(x,x_{o})=y_{o}(x,x_{o})+F(x,x_{o}),\right.
\label{int eq.for y(x)} \\
&&\left. y_{o}(x,x_{o})=-\alpha (x-x_{o}),\right.   \label{eq. yo(x)}
\end{eqnarray}%
where $F(x,x_{o})$ is the solution of the integral equation
\begin{equation}
F(x,x_{o})=\int\limits_{x_{o}}^{x}dx^{\prime }\left( x-x^{\prime }\right)
\left\{ S(x^{\prime },x_{o})-y(x^{\prime },x_{o})^{2}-\frac{2}{x^{\prime }}%
y^{\prime }(x^{\prime },x_{o})\right\} ,  \label{eq.0d}
\end{equation}%
with $S(x^{\prime },x_{o})$ to be expressed in terms of $y(x^{\prime },x_{o})
$ by means of Eq.(\ref{eq.0c}). On the other hand, for finite $x_{o},$ the
Debye-H\"{u}ckel asymptotic approximation for the electrostatic potential
can be proven to hold when measuring the potential at a position $\mathbf{r}$
sufficiently far from the position of the test particle $\mathbf{r}(t).$ In
fact this involves imposing that the so-called \emph{weak-field condition }$%
\widehat{\Phi }_{x_{o}}(x)\ll 1$ be locally satisfied. \ As a consequence,
the normalized electrostatic potential generated by a spherically-symmetric
point particle of charge $Ze$ can be approximated by%
\begin{equation}
\widehat{\Phi }_{x_{o}}(x)\cong \widehat{\Phi }_{x_{o}}^{(ext)}(x)\equiv
\frac{q}{4\pi x}e^{-\Delta x},  \label{eq.1a}
\end{equation}%
where $\Delta x\equiv x-x_{o},$ $q=q(x_{o},\beta )$ is a suitable
dimensionless effective electric charge and $\widehat{\Phi }%
_{x_{o}}^{(ext)}(x)$ is usually known as DH potential. The latter identifies
the \emph{external asymptotic solution} which holds in the subset of the
external domain, outside the Debye sphere, in which the weak-field condition
is satisfied. The remaining notation is standard, thus $\rho =\left\vert
\mathbf{r-r}(t)\right\vert $ is the distance from the point charge and $%
x\equiv \rho /\lambda _{D}$ $\in \left[ x_{o},\infty \right] $ is
the corresponding normalized distance. The weak-field condition is
manifestly locally satisfied both for strongly and weakly coupled
plasmas, provided there results $x>1,$ with $x$ suitably
large (i.e., $x\gg 1$). However, in the case in which there results either $%
x_{o}\ll 1$ and/or $\Gamma \gg 1,$ i.e., for strongly-coupled
plasmas, the electrostatic potential may actually decay on a scale
much shorter than the Debye length and precisely on a scale
suitably close to the boundary of the plasma sheath, i.e., for
$x_{o}\lesssim x\ll 1.$ This conclusion is consistent both with
previous analytic estimates based on the exact solution of the 1D
Debye screening problem \cite{Clemente 1991} and direct numerical
simulations of the 3D Debye screening problem
\cite{Robbins-1988,Dupont1996,Allayrov 1998,Bystrenko 1999}. It
implies, however, that the DH approximation typically holds also
at distances suitably larger than the boundary of the plasma
sheath $x_{o}$ and which extend also to values comparable to the
Debye length, i.e., for $x_{o}\ll x\lesssim 1$
\cite{Tessarotto1992}. Finally, as a further consequence, the
effective electric charge carried by the DH potential $q$ may
appear significantly reduced with respect to the case of weakly
coupled plasmas \cite{Robbins-1988,Tsytovich 1994,Allayrov
1998,Bystrenko 1999,Bystrenko 2003}. To display this effect, it is
convenient to represent the normalized effective charge of the
test particle in the form $q\equiv \beta f_{c}$,  $f_{c}$ being
the \emph{charge reduction factor.} By dimensional analysis, it
follows that $f_{c}$ must be of the form $f_{c}=f_{c}(x_{o},\beta
,\xi _{s}),$ i.e., $f_{c}$ is function, to be
determined, of the only dimensionless parameters of the problem, namely $%
x_{o}$ and $\Gamma $ (or $\beta $) and moreover the ratios $\xi
_{s}$ (with $s=e,i,d$)$.$ \ It is well-known that for
weakly-coupled plasmas $q$ can be approximated by the normalized
electric charge of the isolated test particle, i.e., $q\cong \beta
,$ which implies $f_{c}\cong 1.$ Instead, for strongly-coupled
plasmas,\ the value of charge reduction factor, determined either
via one-dimensional analytic estimates \cite{Clemente 1991} or by
means of a variety of numerical simulation methods (see, for example, \cite%
{Robbins-1988,Dupont1996,Allayrov 1998,Bystrenko 1999}), has been
found to be notably smaller than unity, thus suggesting the
existence of a possible significant \emph{charge reduction
effect}. \ Indeed the investigation of the effective interactions
characterizing high-Z grains in plasmas has attracted interest in
recent years especially for their role in dusty plasmas
\cite{Tsytovich 1994,Bystrenko 2003}. However, for
strongly-coupled plasmas the precise form of the function $f_{c}$
is still unknown and, in particular, an analytic estimate of the
effective charge characterizing the DH potential in
strongly-coupled plasmas is not yet available.

Besides being of obvious interest as a still unsolved mathematical problem,
it is especially important from the physical standpoint to estimate $f_{c}$
as a function of the parameters $x_{o}$ and $\Gamma $ for strongly coupled
plasmas\ characterized either by a high density and/or by the presence of a
large fraction of highly charged particles (grains), such as dusty plasmas
and colloidal suspensions \cite{Pieransky 1983,Thomas 1994}.

In a previous work \cite{Tessarotto2005}, the Debye screening
problem (DSP) has been formulated for an electron-hydrogen plasma
in order to analyze the asymptotic properties of its solutions
near the plasma sheath (i.e., for $x$ suitably close to $x_{o}$).\
For this purpose the case of a strongly-coupled plasma was
investigated in which the parameters $x_{o}$ and $\Gamma $ satisfy
the
asymptotic conditions, to be denoted as \emph{strong-coupling ordering}, $%
\Gamma \sim \frac{1}{O(\delta )}\gg 1$ with $x_{o}\sim O(\delta ^{k})$, $%
k=0,1$and $\delta $ an infinitesimal. The similar case of three-species
plasma can be obtained directly. As a result, invoking for $\widehat{\Phi }%
_{x_{o}}(x)$ the representation (\ref{eq.0c}), together with Eqs.(\ref{int
eq.for y(x)}),(\ref{eq. yo(x)}) and imposing the appropriate boundary
conditions for $\widehat{\Phi }_{x_{o}}(x)$, it follows that the integral on
the r.h.s. of Eq.(\ref{eq.00}) can be estimated asymptotically to yield an
asymptotic equation for the initial condition $\widehat{\Phi }_{o},$ i.e.,
\begin{equation}
\widehat{\Phi }_{o}(x_{o},\beta )\cong \frac{1}{\left\vert Z_{i}\right\vert }%
\ln \left\{ \frac{\left\vert Z_{i}\right\vert }{\xi _{i}}\frac{\Gamma }{%
x_{o}^{2}}\left[ \Gamma -\widehat{\Phi }_{o}\right] \right\} .  \label{eq.5b}
\end{equation}%
In particular, \ for an electron-hydrogen plasma in which the contribution
to $\lambda _{D}$ due to the dusty species is negligible, there results $%
\left\vert Z_{i}\right\vert =1$ and $\xi _{i}=1/2$ and this reduces (\ref%
{eq.5b}) to the expression given in Ref. \cite{Tessarotto2005}. In
addition,\ neglecting higher-order infinitesimals of order $O(\delta ^{r})$ (%
$r>0$ being a suitable real number), the following inequality can
immediately be proven

\begin{eqnarray}
&&\left. f_{c}\leq c^{(a)}(x_{o},\beta )\left[ 1+O(\delta ^{r})\right]
,\right.   \label{eq.5} \\
&&\left. c^{(a)}(x_{o},\beta )\equiv x_{o}\widehat{\Phi }_{o}(x_{o},\beta
),\right.   \label{eq.6}
\end{eqnarray}%
$c^{(a)}(x_{o},\beta )$ being an\emph{\ asymptotic upper bound} for the
charge reduction factor. This result, as indicated below, is relevant to
investigate the charge reduction\ effect in strongly-coupled plasmas.

In this paper we propose a more accurate asymptotic approximation for $%
f_{c}, $ potentially useful to\ analyze the charge reduction effect of
high-Z test particles taking place in strongly-coupled plasmas. In
particular, we intend to prove that, in validity of the previous
strong-coupling ordering, the charge reduction factor can be approximated in
the form:

\begin{equation}
f_{c}(x_{o},\beta )\cong c^{(b)}(x_{o},\beta )\equiv x_{c}e^{\Delta x_{c}}%
\widehat{\Phi }_{x_{o}}^{(int)}(x_{c})/\beta ,  \label{eq.7}
\end{equation}%
where, $x_{c}$ is defined so that $x_{c}>x_{o}$ and is provided by the root
of the equation%
\begin{equation}
y^{\prime }(x_{c},x_{o})=-1-\frac{1}{x_{c}}.  \label{eq.7b}
\end{equation}%
Here $\widehat{\Phi }_{x_{o}}^{(int)}(x)$ denotes the leading-order
contribution to the internal asymptotic solution of the Debye-Poisson
equation (\ref{eq.0}), which can be obtained from the representation (\ref%
{eq.0c}),(\ref{int eq.for y(x)}),(\ref{eq. yo(x)})\ and the integral
equation (\ref{eq.0d}) and holds in an appropriate neighborhood of the
plasma sheath ($x_{o}$) \cite{Tessarotto2005}. There results%
\begin{equation}
\widehat{\Phi }_{x_{o}}^{(int)}(x)\cong \exp \left\{ y_{o}(x,x_{o})\right\} ,
\label{eq.9}
\end{equation}%
while $y^{\prime }(x,x_{o})$ reads%
\begin{eqnarray}
&&\left. y^{\prime }(x,x_{o})\cong -\alpha -\alpha ^{2}\Delta x+2\alpha \log
\frac{x}{x_{o}}+\right.  \\
&&\left. +\frac{\xi _{i}}{\widehat{\Phi }_{o}}\int\limits_{x_{o}}^{x}dx^{%
\prime }\exp \left\{ \left\vert Z_{i}\right\vert \widehat{\Phi }_{o}\exp
\left\{ -\alpha \Delta x^{\prime }\right\} +\alpha \Delta x^{\prime }+\frac{1%
}{2}\alpha ^{2}\Delta x^{\prime 2}-2\alpha \left[ x^{\prime }\log \frac{%
x^{\prime }}{x_{o}}-\Delta x^{\prime }\right] \right\} ,\right.   \notag
\end{eqnarray}
where $\Delta x^{\prime }=x^{\prime }-x_{o}.$ Eqs.(\ref{eq.7}) and (\ref%
{eq.7b}) uniquely follow by imposing suitable matching conditions
between the internal and external asymptotic solutions, $\widehat{\Phi }%
_{x_{o}}^{(ext)}(x)$ and $\widehat{\Phi }_{x_{o}}^{(int)}(x),$\ precisely%
\begin{eqnarray}
\widehat{\Phi }_{x_{o}}^{(int)}(x_{c}) &=&\widehat{\Phi }%
_{x_{o}}^{(ext)}(x_{c}),  \label{eq.10} \\
\left. \frac{\partial }{\partial x}\widehat{\Phi }_{x_{o}}^{(int)}(x)\right%
\vert _{x=x_{c}} &=&\left. \frac{\partial }{\partial x}\widehat{\Phi }%
_{x_{o}}^{(ext)}(x)\right\vert _{x=x_{c}}.  \label{eq.11}
\end{eqnarray}%
It is immediate to prove that Eq.(\ref{eq.11}) is equivalent to the
condition of stationarity for the function $f_{c}^{(a)}(x,x_{o},\beta )$
\begin{equation}
\left. \frac{\partial }{\partial x}f_{c}^{(a)}(x,x_{o},\beta )\right\vert
_{x=x_{c}}=0.  \label{eq.8}
\end{equation}%
Therefore Eq.(\ref{eq.8}) also furnishes the same asymptotic approximation
for the charge reduction factor given by Eq.(\ref{eq.7}). The accuracy of
the asymptotic approximation $c^{(b)}(x_{o},\beta )$ and of the upper bound $%
c^{(a)}(x_{o},\beta )$ can be tested by comparing them with the estimate of
the charge reduction factor obtained from direct numerical simulations of
the 3D Debye screening problem. For this purpose the function

\begin{equation}
g(x,x_{o},\beta )=xe^{x-x_{o}}\widehat{\Phi }_{x_{o}}(x)/\beta  \label{eq.12}
\end{equation}%
has to be evaluated numerically by solving the Debye-Poisson equation. The
charge reduction factor is defined\ as the limit%
\begin{equation}
f_{c}(x_{o},\beta )=\lim_{x\rightarrow +\infty }xe^{x-x_{o}}\widehat{\Phi }%
_{x_{o}}(x)/\beta .  \label{eq.13}
\end{equation}%
In practice, for numerical estimates it can be approximated by $%
f_{c}(x_{o},\beta )\cong g(x_{f},x_{o},\beta )$ with $x_{f}$ to be suitably
defined. It is important to stress\ that, in the limit $\delta \rightarrow
0, $ one expects $x_{c}\rightarrow 0,$ $x_{f}\rightarrow 0$ and moreover
\begin{equation}
\lim_{\delta \rightarrow 0}c^{(b)}(x_{o},\beta )=\lim_{\delta \rightarrow
0}f_{c}(x_{o},\beta ).  \label{eq.14}
\end{equation}%
In addition, it must result also%
\begin{equation}
\lim_{\delta \rightarrow 0}c^{(b)}(x_{o},\beta )=\lim_{\delta \rightarrow
0}c^{(a)}(x_{o},\beta ).  \label{eq.15}
\end{equation}%
Therefore, for $\delta $ small enough (i.e., for $\Gamma \gg 1$ and/or $%
x_{o}\ll 1$), it it actually possible to approximate $f_{c}(x_{o},\beta )$
both in terms of $c^{(b)}(x_{o},\beta )$ and $c^{(a)}(x_{o},\beta ).$
Numerical simulations indicate, however, that the asymptotic estimate $%
c^{(b)}(x_{o},\beta )$ actually converges rapidly to $f_{c}(x_{o},\beta )$
and furnishes therefore a more accurate approximation than the upper bound $%
c^{(a)}(x_{o},\beta ).$ Nevertheless, particularly thanks to Eq.(\ref{eq.15}%
), $c^{(a)}(x_{o},\beta )$\ can still be used for order-of-magnitude
estimates.\

\begin{figure}[tbh]
\begin{center}
\includegraphics[width=.7\textwidth]{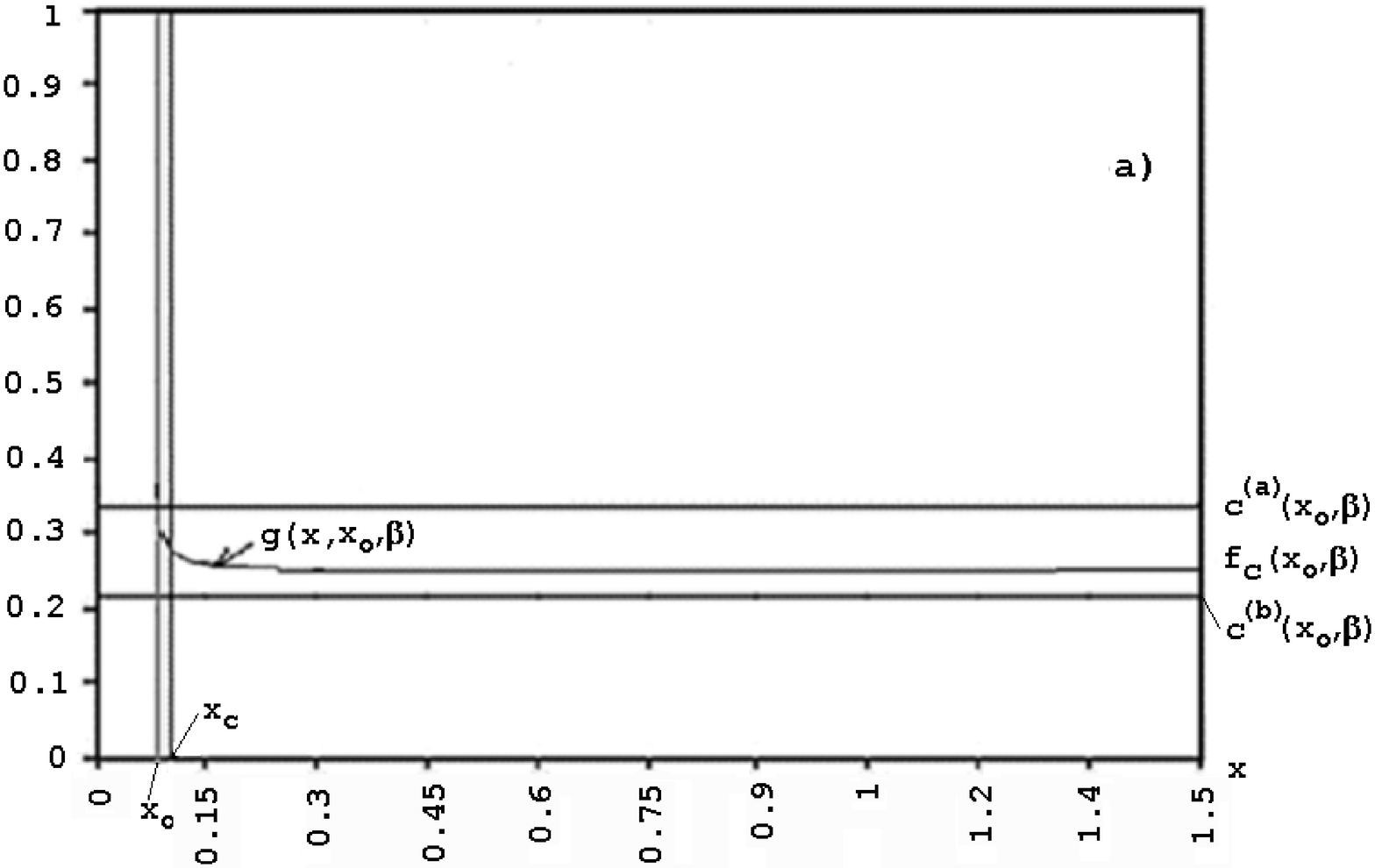} \includegraphics[width=.7%
\textwidth]{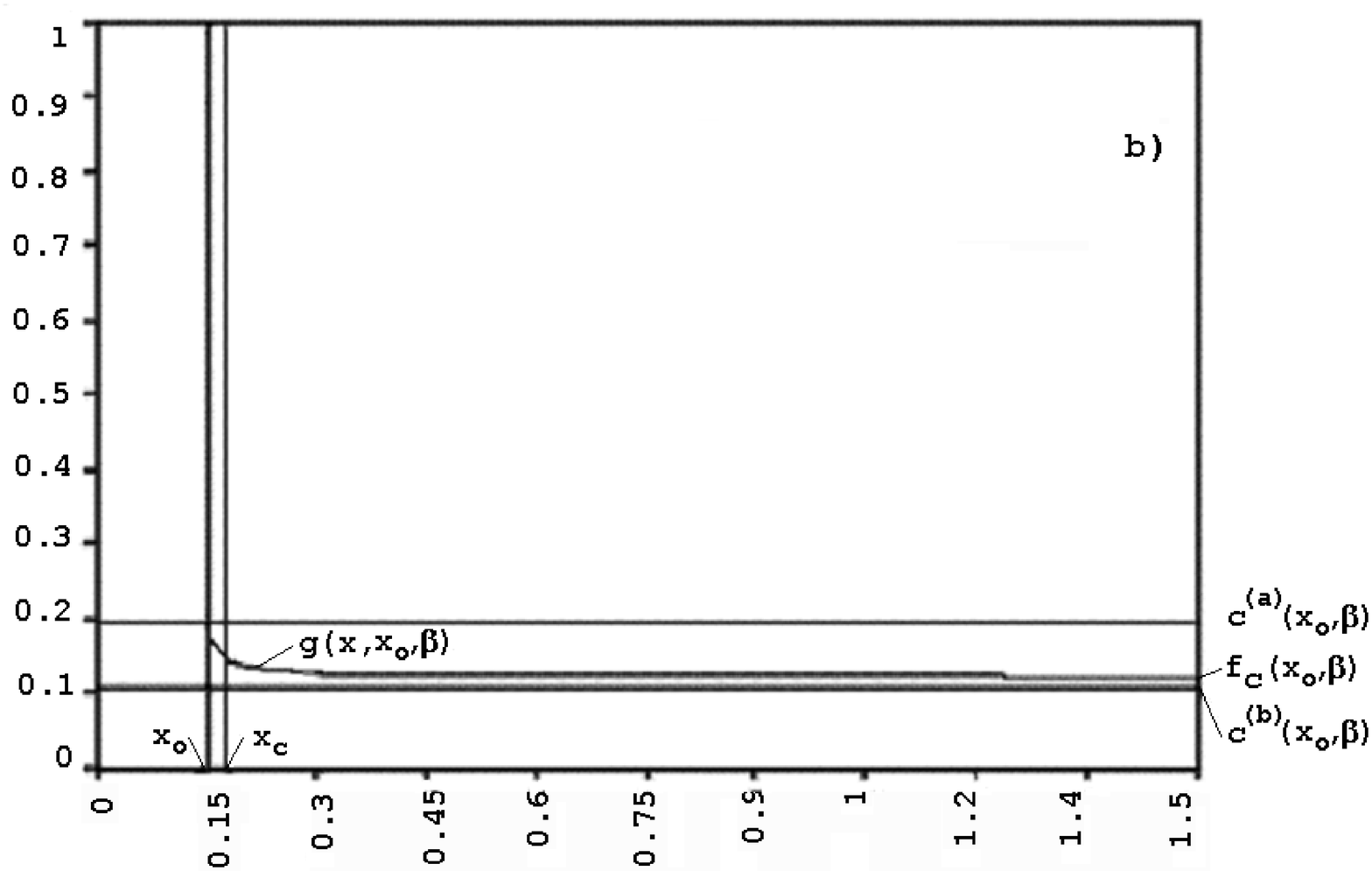}
\end{center}
\par
\vspace{-10pt}
\caption{Comparison between the asymptotic approximation of the charge
reduction factor $c^{(a)}(x_{o},\protect\beta)$, the numerical estimate $%
f_{c}(x_{o},\protect\beta)$ and the upper bound estimate $c^{(a)}(x_{o},%
\protect\beta)$. Case a) corresponds to $x_{o}=0.08$ and $\protect\beta =2.5$
for which there results $x_{c}\protect\cong 0.098$, $c^{(b)}(x_{o},\protect%
\beta)\protect\cong 0.211$, $f_{c}(x_{o},\protect\beta)\protect\cong 0.246$
and $c^{(a)}(x_{o},\protect\beta)\protect\cong 0.33$. Case b) corresponds to
$x_{o}=0.15$ and $\protect\beta =10$ for which there results $x_{c}\protect%
\cong 0.171$, $c^{(b)}(x_{o},\protect\beta)\protect\cong 0.106$, $%
f_{c}(x_{o},\protect\beta)\protect\cong 0.119 $ and $c^{(a)}(x_{o},\protect%
\beta)\protect\cong 0.19$. }
\label{spp:dist0}
\end{figure}

\begin{figure}[tbh]
\begin{center}
\includegraphics[width=.7\textwidth]{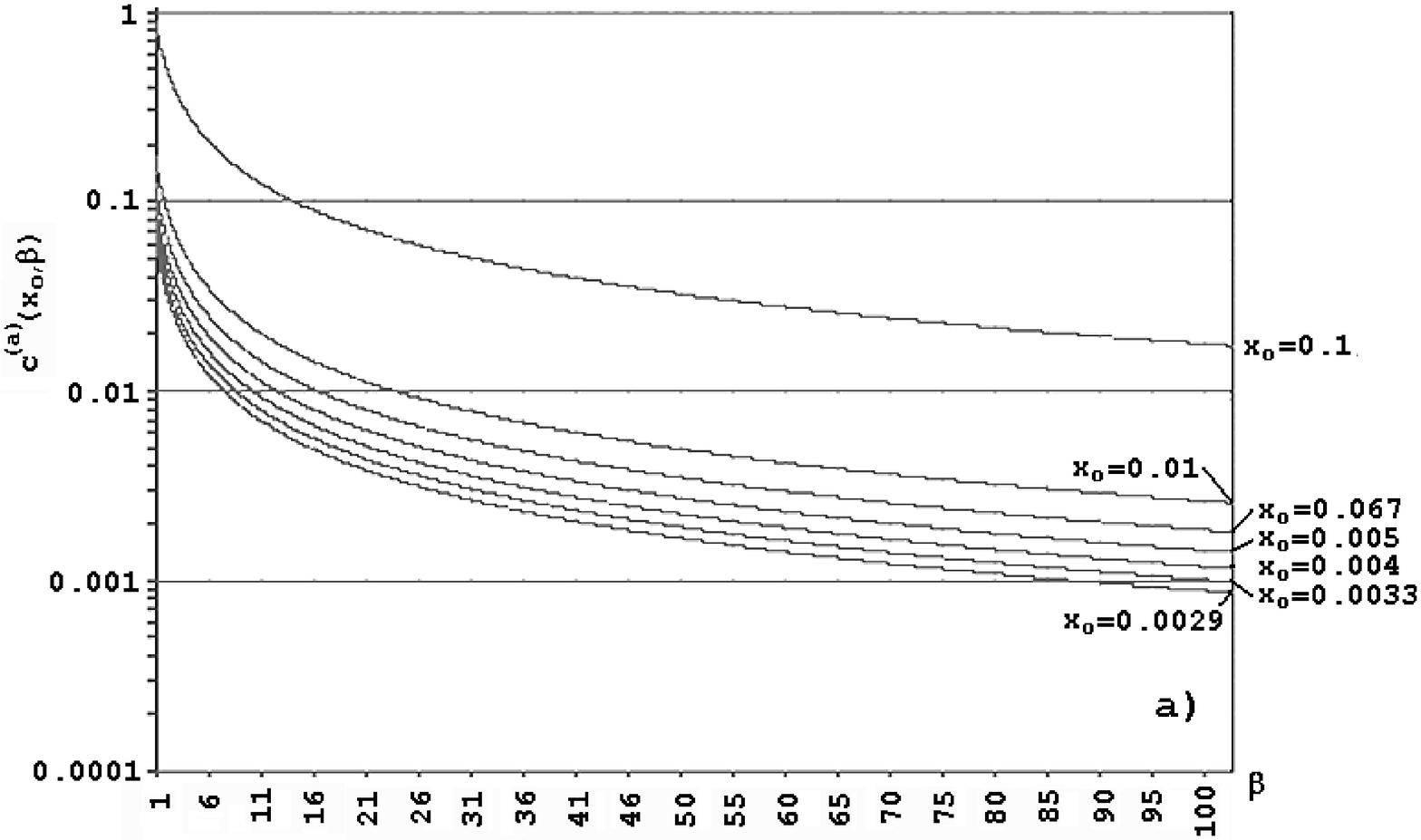} \hspace{-10pt} %
\includegraphics[width=.7\textwidth]{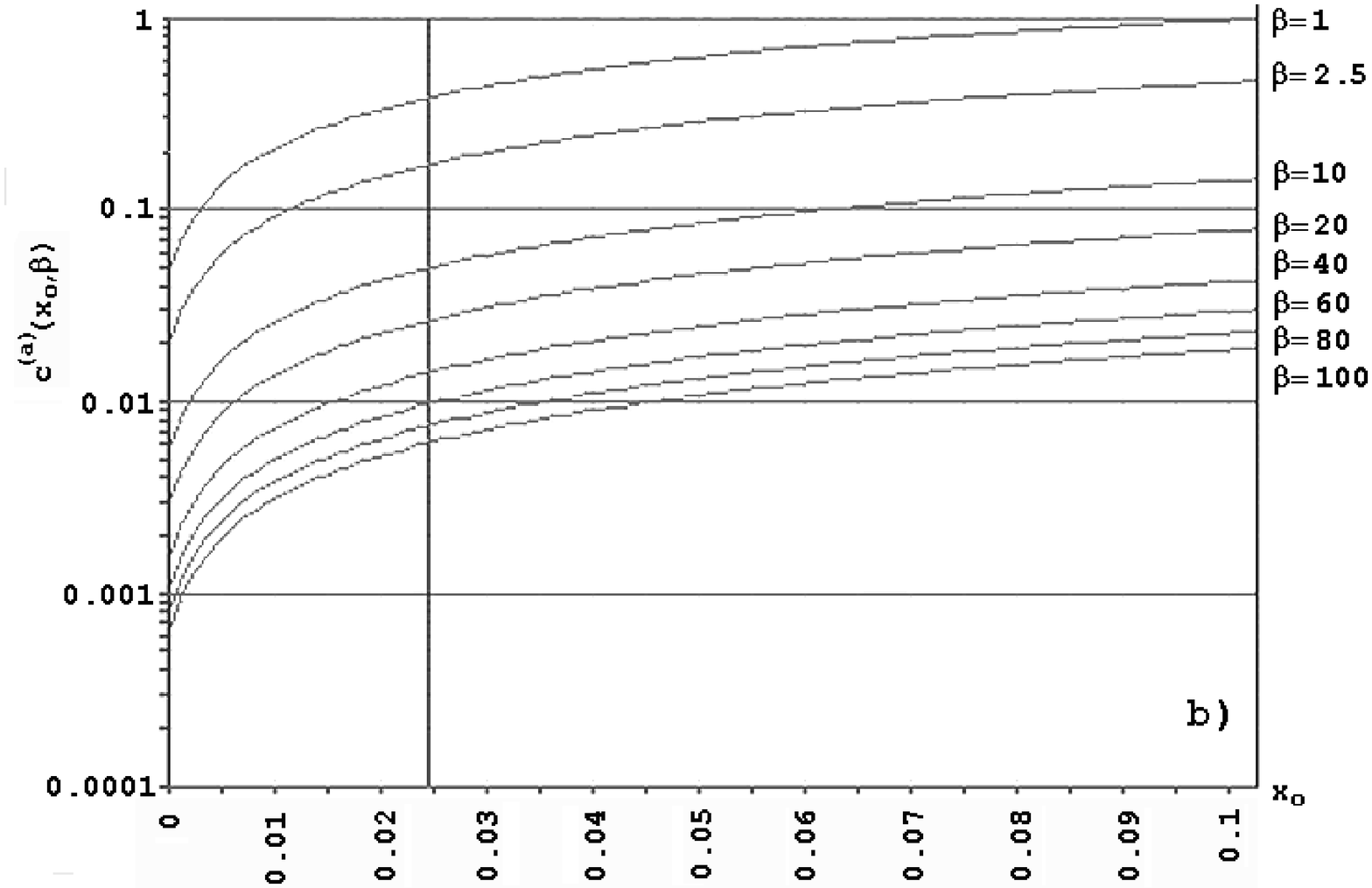}
\end{center}
\par
\par
\vspace{-10pt}
\caption{Plot a) shows $c^{(a)}(x_{o},\Gamma )$ as a function of $\protect%
\beta $ with $x_{o}=0.1,0.06,0.024,0.048)$. Plot b) shows $%
c^{(a)}(x_{o},\Gamma )$ as a function $x_{o}$ for $\protect\beta %
=10,20,40,60,80,100$. }
\label{spp:dist1}
\end{figure}

In figures \ref{spp:dist0} the result of two numerical simulations are
reported, for the case of an electron-hydrogen plasma and for $\xi _{i}=1/2,$
which correspond respectively to $(x_{o},\beta )=(0.08,2.5)$ and $%
(x_{o},\beta )=(0.15,10).$ In both cases the matching point $x_{c}$ is found
to be close to the boundary of the plasma sheath, which implies that the DH
potential actually approximately applies for $x\geq x_{c}.$ The upper bound
estimate $c^{(a)}(x_{o},\beta )$ can be used for order-of-magnitude
estimates of the charge reduction effect. Finally, the parametric
dependencies of $c^{(a)}(x_{o},\beta )$ with respect to $\beta $ and $x_{o}$
are displayed for selected values of the parameters. Figures \ref{spp:dist1}
demonstrate that for typical values of the Coulomb parameter $\Gamma $ and
the radius of the plasma sheath $x_{o}$ the effective electric charge of the
DH potentially appear dramatically reduced.

The present results are relevant for the investigation of
strongly-coupled plasmas, such as dusty plasmas, in which a
significant fraction of high-$Z$ charged particles is present. In
fact, the charge reduction effect has fundamental consequences on
the Debye screening phenomenon, since it actually limits the
magnitude of the Coulomb interactions produced by these particles.
The estimate of the charge reduction factor $f_{c},$ here
obtained, is a necessary prerequisite for the description of a
variety of kinetic processes (for example, condensation
\cite{Bellan2004}, collisional processes \cite{Tsytovich
1999,Tsytovich 2001a}, etc.), as well as electrostatic or
electromagnetic instabilities in dusty plasmas (see for example
\cite{Salimullah 1999,Shukla 2004}). \

\section*{Acknowledgments}

Work supported by PRIN Research Program \textquotedblleft \textit{Programma
Cofin 2004: Modelli della teoria cinetica matematica nello studio dei
sistemi complessi nelle scienze applicate}\textquotedblright ( MIUR Italian
Ministry), the ICTP/TRIL Program (ICTP, Trieste, Italy), and the Consortium
for Magnetofluid Dynamics, University of Trieste, Italy.

\newpage

\pagebreak

\end{document}